\documentclass{article}
\usepackage{amsfonts,amssymb, amsmath}
\usepackage[english]{babel}
\usepackage{graphicx}
\usepackage{float}

\newtheorem{prop}{Proposition}

\newtheorem{ex}{Example}

\def\nn{\nonumber }
\def\bq{ \begin{equation}}
\def\eq{ \end{equation}}
\def\ben{ \begin{eqnarray}}
\def\en{ \end{eqnarray}}
\def\a{{\alpha}}
\def\b{{\beta}}

\begin{document}


\title{Elliptic curve arithmetic and superintegrable systems}

\author{A.V. Tsiganov \\
\it\small St.Petersburg State University, St.Petersburg, Russia\\
\it\small e--mail: andrey.tsiganov@gmail.com}
\date{}
\maketitle

\begin{abstract}
Harmonic oscillator and the Kepler problem are superintegrable systems which admit more integrals  of motion than degrees of freedom and all these integrals are polynomials in momenta. We present superintegrable deformations of the oscillator and the Kepler problem with algebraic and rational first integrals. Also, we discuss  a family of superintegrable metrics on the two-dimensional sphere, which have similar first integrals.
 \end{abstract}

\section{Introduction}
\setcounter{equation}{0}
In 1757-1759 Euler created the theory of elliptic integrals which in turn gave birth to the Abel theory of Abelian integrals, to the Jacobi theory of elliptic functions, to the Riemann theory of algebraic functions, etc. This paper considers two themes in algebraic geometry and elliptic curve cryptography  descended from Euler’s work: elliptic curve arithmetic and algebraic integrals of Abel's equations, see Problems 81-84 in Euler's textbook \cite{eul0}.

In 1760-1767 Euler applied this mathematical theory to searching of algebraic trajectories in the two fixed centers problem. In \cite{eul1,eul2,eul3} he reduced equations of motion to one equation defining trajectory
\[
\dfrac{dr}{\sqrt{R}}+ \dfrac{ds}{\sqrt{S}}=0\,,
\]
 identified algebraic integral  of this equation with a partial first integral in the phase space  and separated partial algebraic trajectories from transcendental trajectories.  In particular, Euler obtained an additional first integral for the superintegrable  Kepler problem, which is a partial case of two fixed centers problem,  in terms of elliptic coordinates on the plane.  In \cite{lag1} Lagrange proved that equations of motion for the two fixed centers problem with three degrees of freedom are separable in prolate spheroidal coordinates,  considered generalized two fixed centers problem and then used algebraic integral of Abel's equation  for searching algebraic trajectories in this generalized two fixed centers problem, see \cite{lag1,lag2} and comments by Serret \cite{ser} and Darboux \cite{darb}. Modern description of algebraic trajectories in the two centers problem  may be found in \cite{dul16}.

Thus, if generic or partial equations of motion are reduced to  Abel's equation on the elliptic curve, then we have additional partial or complete first integral obtained by Euler in his solution of  Problems 81-84 in \cite{eul0}.  In  \cite{ts09a,ts08,ts09,ts10} we used Euler's construction  in order to classify known superintegrable systems with additional first integrals which are polynomials in  momenta.
This paper considers superintegrable deformations of the Kepler problem, harmonic oscillators on the plane  and geodesics on the sphere which have algebraic and rational additional integrals of motion.

\subsection{Arithmetic of elliptic curves }
Let us consider smooth nonsingular elliptic  curve $X$ on the projective plane defined by an equation of the form
\bq\label{ell-curve}
X:\quad y^2=f(x)\,,\qquad f(x)=a_4x^4+a_3x^3+a_2x^2+a_1x+a_0\,.
\eq
The prime divisors are points on $X$,  denoted $P_i = (x_i, y_i) $,   including  point at infinity $P_\infty$, which plays  the role of neutral element $0$ in arithmetic of elliptic curves.

In 1757 Euler proved an addition formulae for elliptic integrals, in modern terms he proved that by adding
two points on $X$
\[ (x_1, y_1) + (x_2, y_2) = (x_3, y_3) \]
one gets the third point with the following abscissa and ordinate
\bq\label{add-gen}
x_3=-x_1-x_2-\dfrac{2b_0b_2+b_1^2-a_2}{2b_1b_2-a_3}\,,\qquad\mbox{and}\qquad y_3=-\mathcal P(x_3)\,,
\eq
where
\[
\mathcal P(x)=b_2x^2+b_1x+b_0=\sqrt{a_4}(x-x_1)(x-x_2)+\dfrac{(x-x_2)y_1}{x_1-x_2}+\dfrac{(x-x_1)y_2}{x_2-x_1}.
\]
Then Euler explicitly defined doubling of divisor
\[[2]P_1=(x_1,y_1)+(x_1,y_1)=([2]x_1,2[y_1]),\]
 i.e. point of $X$ with  coordinates
\bq\label{doub-ell}
 \begin{array}{rcl}
[2]x_1&=&-2x_1-\dfrac{2b_0b_2+b_1^2-a_2}{2b_1b_2-a_3}\,,\qquad [2]y_1=-\mathcal P\bigl([2]x_1\bigr)\,,\\
\\
\mathcal P(x)&=&b_2x^2+b_1x+b_0=\sqrt{a_4}(x-x_1)^2+\dfrac{(x-x_1)(4a_4x_1^3+3a_3x_1^2+2a_2x_1+a_1)}{2y_1}+y_1\,,
\end{array}
\eq
and tripling of divisor
\[[3]P_1=([2]x_1,2[y_1])+(x_1,y_1)=([3]x_1,3[y_1])\,,\]
i.e. point of $X$ with coordinates
\bq\label{trip-ell}
\begin{array}{rcl}
[3]x_1&=& -3x_1-\dfrac{a_3-2b_1b_2}{a_4-b_2^2}\,,\qquad [3]y_1=-\mathcal P\bigl([3]x_1\bigr)\,,\\
\\
\mathcal P(x)&=&b_2x^2+b_1x+b_0=-\dfrac{(x-x_1)^2 (4 a_4 x_1^3+3 a_3 x_1^2+2 a_2 x_1+a_1)^2}{8 y_1^3}\\
\\
&+&\dfrac{(x-x_1)\Bigl(x \bigl(6 a_4 x_1^2+3 a_3 x_1+a_2\bigr)-2 a_4 x_1^3+a_2 x_1+a_1\Bigr)}{2 y_1}+y_1\,,
\end{array}
\eq
and described an algorithm for multiplication on any integer $m$, see Problem 83 in \cite{eul0}.  Later Abel used  elliptic curve point multiplication in proving his theorem on  $m$-division points of the lemniscate when he introduced some pre-image of the division polynomials. Modern computer algorithms for performing addition and multiplication on elliptic curve are discussed in \cite{blake99,hand06,was08}.

Lagrange proves Euler's addition equation introducing time $t$ and  motion of two points $P_{1}(t)$ and $P_{2}(t)$ on $X$ governed by the following Newton equations
\[
\dfrac{d^2x_1}{dt^2}=2y_1^2\,,\qquad \dfrac{d^2x_2}{dt^2}=2y_2^2\,,
\]
see details in \cite{gr}, p.144.  In this terms  Euler’s solutions of Problems 81-84 \cite{eul0}  can be reformulated in the following form:
 If two points $P_1=(x_1,y_1)$ and $P_2(x_2,y_2)$ move along  a  curve $y^2=f(x)$,  there  is  an  algebraic  constraint  on their motion with the property that
 \[
 \int \omega(x_1,y_1)dx_1+\int \omega(x_2,y_2)dx_2
 \]
 can  be  expressed,   for  any  differential $\omega(x,y)$,  in  terms  of elementary functions in  coordinates $x_1,y_1$ and $x_2,y_2$ when these coordinates satisfy the algebraic constraint. In his famous Paris memoir \cite{ab}, Abel states Euler's conclusion
in almost exactly this form as a preamble to his famous theorem.

In Problem 81 Euler calculates algebraic constraint associated with addition of points and  proves  that
\bq\label{eul-int}
C=\left(\dfrac{y_1-y_2}{x_1-x_2} \right)^2-a_4(x_1+x_2)^2-a_3(x_1+x_2)
\eq
is the general integral of the differential relation
\bq\label{eul-eq}
\dfrac{dx_1}{y_1}+\dfrac{dx_2}{y_2}=0
\eq
when $C$ is a constant and particular integral of
\[
\dfrac{dx_1}{y_1}+\dfrac{dx_2}{y_2}+\dfrac{dx_3}{y_3}=0
\]
when
\bq\label{eul-int-4}
C=2a_4x_3^2+a_3x_3+a_2-2\sqrt{a_4}y_3\,.
\eq
In 1863 Clebsch proposed geometric approach to  construction of algebraic constraints, closely interwoven with the intersection theory,  which was continued by Brill and Noether in 1857 and formalized  by Poincar\'{e} in 1901 and Severy in 1914, see classical textbooks \cite{gr,hens}, review \cite{bliss24}  and modern discussion in  \cite{cost12}.

Then in Problem 83 Euler proves that
\bq\label{eul-int-nm}
 C_{mn}=\left(\dfrac{[m]y_1-[n]y_2}{[m]x_1-[n]x_2} \right)^2-a_4\bigl([m]x_1+[n]x_2\bigr)^2-a_3\bigl([m]x_1+[n]x_2\bigr)
 \eq
 is the general integral of the differential relation
\bq\label{eul-eq-nm}
m\dfrac{dx_1}{y_1}+n\dfrac{dx_2}{y_2}=0\,,
\eq
associated with scalar multiplication of points on integer numbers $m,n$ and addition of the obtained results. Here and below we write the coordinates of $[m](x, y)$ as $([m]x,[m[y])$.

In fact Euler proposed only an algorithm of computations, because explicit expression for $C_{mn}$ is a cumbersome  formula.
For instance, we have
\bq\label{eul-int-21}
C_{21}=\dfrac{A}{B^2}=\frac{16A_6y_1^6+32A_5y_1^5+16A_4y_1^4+16{f'_1}A_3y_1^3+4{f'_1}^2A_2y_1^2+4{f'_1}^3A_1y_1+{f'_1}^4A_0}
{\left(B_3y_1^3+B_2y_1^2+B_1y_1+B_0\right)^2}
\eq
where $f'_1=df(x_1)/dx_1$ is a derivative of the polynomial $f(x)$ from (\ref{ell-curve}) at  point $x=x_1$,
\[
B= 8\sqrt{a_4}y_1^3-4\bigl(2a_4x_1(x_1+2x_2)+a_3(2x_1+x_2)+a_2\bigr)y_1^2-4\sqrt{a_4}(x_1-x_2)f'_1y_1+{f'_1}^2
\]
and
\[\begin{array}{rcl}
A_0&=&\scriptstyle  2a_4x_2^2+a_3x_2+a_2-2\sqrt{a_4}y_2\,,\\ \\
A_1&=&\scriptstyle -4a_4^{3/2}x_1x_2^2-a_2^{1/2}(2a_3x_1x_2+a_3x_2^2+2a_2x_1+a_1)+(4a_4x_1+a_3)y_2\,,
\\ \\
A_2&=&\scriptstyle -4(x_1^4+x_2^4)a_4^2-4\bigl(a_3(x_1^3+x_2^3)+a_2(x_1^2+x_2^2)-a_1x_1\bigr)a_4
+a_1a_3\\
&&\scriptstyle  -2a_2^2-2a_3(2x_1+x_2)a_2-x_2(4x_1-x_2)a_3^2+4a_4y_2^2\,,
\\ \\
A_3&=&\scriptstyle 8x_1(x_1^2+x_2^2)^2 a_4^{5/2}
+\bigl(2a_3(x_1^2+x_2^2)(5x_1^2+4x_1x_2+x_2^2)+8a_2x_1(2x_1^2+x_2^2)+3a_1(2x_1^2+x_2^2)\bigr)a_4^{3/2}\\
&&\scriptstyle +\bigl(
a_3^2(2x_1^3+7x_1^2x_2+x_1x_2^2+2x_2^3)+4a_2^2x_1+a_1a_3(3x_1+x_2)+a_2a_3(9x_1^2+4x_1x_2+x_2^2)+2a_1a_2
\bigr)a_4^{1/2}
\\
&&\scriptstyle -\bigl(16a_4^2x_1^3+2(6a_3x_1^2-a_1)a_4+a_3(3a_3x_1+a_2)\bigr)y_2-2a_4^{1/2}(4a_4x_1+a_3)y_2^2\,,
\end{array}
\]
\[\begin{array}{rcl}
A_4&=&\scriptstyle -8x_1^2x_2^2(3x_1^2+2x_2^2)a_4^3-4x_1
\bigl(3a_1(2x_1^2+x_2^2)+a_2x_1(3x_1^2+2x_2^2)+a_3x_2(3x_1^3+6x_1^2x_2+4x_1x_2^2+2x_2^3)\bigr)a_4^2
\\
&&\scriptstyle -\bigl(
a_1^2+8a_1a_2x_1+a_1a_3(18x_1^2+4x_1x_2+2x_2^2)-2a_2^2x_2^2+4a_2a_3x_1^2x_2
+a_3^2x_2(4x_1^3+8x_1^2x_2+8x_1x_2^2+x_2^3)\bigr)a_4
 \\
 &&\scriptstyle
-a_1(3a_3^2x_1+a_2a_3)+a_2^3+a_2^2a_3(4x_1+x_2)+a_2a_3^2(3x_1^2+4x_1x_2-x_2^2)-a_3^2(x_1-x_2)(x_1^2-3x_1x_2-x_2^2)
 \\
&&\scriptstyle +2a_4^{1/2}\bigl(20a_4^2x_1^4+4x_1(5a_3x_1^2+a_2x_1-a_1)a_4+6a_3^2x_1^2+4a_2a_3x_1-a_1a_3+a_2^2\bigr)
y_2\\
&&\scriptstyle +(16a_4^2x_1^2+8a_4a_3x_1+a_3^2)y_2^2\,,\\
\\
A_5&=&\scriptstyle -
8a_4^{5/2}x_1^2(2x_1^2+x_2^2)+4a_4^{3/2}(4a_3x_1^3+a_3x_1^2x_2+a_3x_1x_2^2+3a_2x_1^2+a_2x_2^2)\\
&&\scriptstyle +a_4^{1/2}\bigl(3a_3^2x_1^2+4a_3^2x_1x_2-a_3^2x_2^2+6a_2a_3x_1+2a_2a_3x_2+2a_2^2\bigr)
-\bigl(8a_4^2x_1^2+4a_4(a_3x_1+a_2)-a_3^2\bigr)y_2\,,\\
\\
A_6&=&\scriptstyle  8(2x_1^2+x_2^2)a_4^2+4(2a_3x_1+a_3x_2+a_2)a_4^2+a_3^2+8a_4^{3/2}y_2\,.
\end{array}
\]
If $m=3$ and $n=1$, then  one gets an algebraic integral with  similar structure
\bq\label{eul-int-31}
C_{31}=\dfrac{A}{B^2}=\frac{A_{14}y_1^{14}+A_{13}y_1^{13}+\ldots+A_0}{\left(B_6y_1^6+B^4y_1^4+B_2y_1^2+B_0\right)^2}\,,
\eq
where  $f''_1=d^2f(x_1)/dx_1^2$ and
\[\begin{array}{l}
B=
 64(3a_4x_1+a_4x_2+a_3)y_1^6-8f''_1\bigl(2a_4x_1^2(x_1+3x_2)+3a_3x_1(x_1+x_2)+a_2(3x_1+x_2)+2a_1\bigr)y_1^4
\\
\\
+8{f'_1}^2\bigl(2a_4x_1^2(3x_2-x_1)+3a_3x_1x_2+a_2(x_1+x_2)+a_1\bigr)y_1^2
+{f'_1}^4(x_1-x_2)\,.
\end{array}
\]
For brevity we omit expressions for  functions $A_k$, which are polynomials in  coordinates $x_{1,2} $ and $y_2$. They can be obtained using any computer algebra system.

In the next Section we apply these Euler's results to construction of rational and algebraic functions commuting with Hamilton functions of superintegrable systems with two degrees of freedom.

\section{Superintegrable systems with two degrees of freedom}
The traditional way of writing an elliptic curve equation of  is to use its short or long Weierstrass form.  In elliptic curve cryptography we can find also other forms of the elliptic curve such as  Edwards  curves,  Jacobi  intersections  and  Jacobi quartics, Hessian curves,  Huff curves, etc.

Following \cite{lang78} we begin with nonsingular elliptic curve $X$  defined by a short  Weierstrass equation
\bq\label{wei-eq}
 y^2=f(x)\,,\qquad f(x)= x^3+ax+b\,,
\eq
 and arithmetic equation on $X$
\[
P_1+P_2+P_3=0\,.
\]
 Here $P_1,P_2$ and $P_3$ are intersection points of $X$ with a straight line, see standard picture in Figure 1 and in textbooks \cite{blake99,hand06,was08}.
  \begin{figure}[!ht]
\center{\includegraphics[width=0.7\linewidth, height=0.2\textheight]{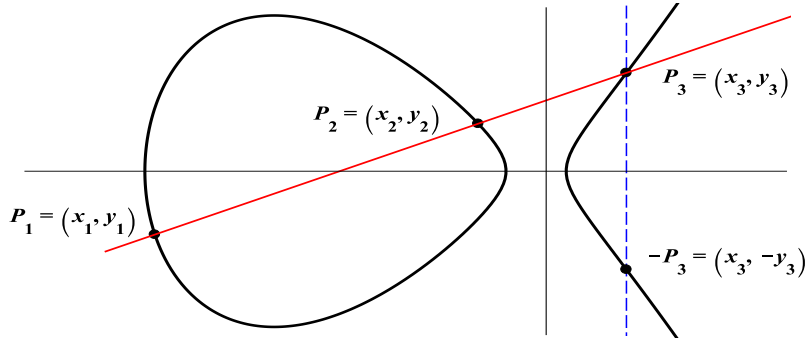} }
\caption{ Addition of points $P_1+P_2+P_3=0$ on the elliptic curve.}
\end{figure}
\par\noindent
Using  coordinates of  points $P_1=(x_1,y_1)$ and $P_2(x_2,y_2)$
 we can easily define  coordinates of the third point
\[
x_3=\lambda^2-(x_1+x_2)\,,\qquad y_3=y_1+\lambda(x_3-x_1)\,,\qquad
                                                                            \lambda=\frac{y_2-y_1}{x_2-x_1}\,.
\]
In this case  Euler's integral (\ref{eul-int})  coincides with  abscissa of $P_3$, i.e. $C=x_3$ and it is the required algebraic constraint for motion of two points $P_1(t)$ and $P_2(t)$ along the elliptic curve $X$.

Let us also present a well-known expression for the elliptic curve point multiplication on any positive integer $m$:
  \[
 [m](x,y)\equiv\bigl[m]x,[m]y)=\left(x-\dfrac{\psi_{m-1}\psi_{m+1}}{\psi_m^2}\,,\dfrac{\psi_{2m}}{2\psi_m^4} \right)
 \]
 where $\psi_m$ are the so-called division polynomials in $\mathbb Z[x,y,a,b]$, which are the ratio of two  Weierstrass
$\sigma$-functions, see \cite{blake99,lang78,was08}. It is easy to see that  abscissa of $[m]P$ is a rational function strictly in terms
of $x$ whereas ordinate has the form $yR(x)$, where $R(x)$ is a rational function.

 If we identify abscissas and ordinates of  points $P_1$ and $P_2$ with canonical coordinates on the phase space
\[
x_{1,2}=u_{1,2}\,,\qquad y_{1,2}=p_{u_{1,2}}\,,
\]
where
\[
\{u_1,u_2\}=0\,,\quad \{p_{u_1},p_{u_2}\}=0\,,\quad
\{u_i,p_{u_j}\}=\delta_{ij}\,,
\]
and solve a pair of equations $y_i^2=f(x_i)$\,, $i=1,2$\,, with respect to $a,b$, we obtain two functions on the phase space
\bq\label{wei-int}
a=\dfrac{p_{u_1}^2}{u_1-u_2}+\dfrac{p_{u_2}^2}{u_2-u_1}-u_1^2-u_1u_2-u_2^2\,,\qquad
b=\dfrac{u_2p_{u_1}^2}{u_2-u_1}+\dfrac{u_1p_{u_2^2}}{u_1-u_2}+(u_1+u_2)u_1u_2\,,
\eq
which are in involution with respect to canonical Poisson brackets.

Taking  $H=a$ as a Hamiltonian, one gets integrable system on the phase space $T^*\mathbb R^2$ with quadratures
\[
\omega_1=\int \dfrac{u_1du_1}{\sqrt{u_1^3+au_1+b}}+
\int \dfrac{u_2du_2}{\sqrt{u_2^3+au_2+b}}=-2t
\]
and
\[
\omega_2=\int \dfrac{du_1}{\sqrt{u_1^3+au_1+b}}+
\int \dfrac{du_2}{\sqrt{u_2^3+au_2+b}}=\mathrm{const}\,.
\]
 In physical  terms $a,b$ and  $\omega_{1},\omega_2$ are action-angle variables associated with this motion, whereas  Euler's algebraic  constraint  (\ref{eul-int}) is an additional first integral \cite{ts18a}.

Second quadrature in the differential form coincides with (\ref{eul-eq}) and, therefore,  we have superintegrable system with additional first integrals which are abscissa and ordinate of the  third point $P_3$ on a projective plane
\[
x_3=\left(\dfrac{p_{u_1}-p_{u_2}}{u_1-u_2}\right)^2-(u_1+u_2)\,,\qquad
y_3=p_{u_1}+\left(\dfrac{p_{u_1}-p_{u_2}}{u_1-u_2}\right)(x_3-u_1)\,.
\]
Functions $a, b$ from (\ref{wei-int}) and functions $x_3, y_3$ on the phase space  $T^*\mathbb R^2$ form an algebra of integrals
\[\begin{array}{lll}
 \{a,b\}=0\,,\qquad &\{a,x_3\}=0\,,\qquad &\{a,y_3\}=0\,,\\ \\
 \{b,x_3\}=2y_3\,,\qquad &\{b,y_3\}=3x_3^2+a\,,\qquad &\{x_3,y_3\}=-1\,,
 \end{array}
\]
in which Weierstrass equation (\ref{wei-eq}) plays the role of syzygy
\[
y_3^2=x_3^3+ax_3+b\,.
\]
In this case two points $P_1(t)$ and $P_2(t)$ move along  a  curve $y^2=f(x)$ with fixed third point $P_3$ because its abscissa $x_3$ and ordinate $y_3$ are additional first integrals,  see the picture in Figure 2.
  \begin{figure}[!ht]
\center{\includegraphics[width=0.7\linewidth, height=0.2\textheight]{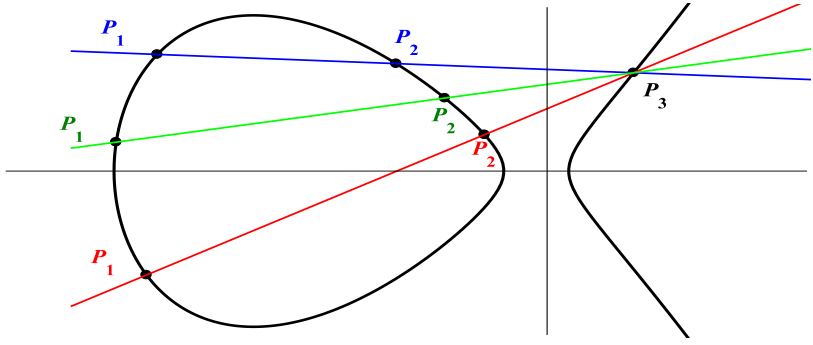} }
\caption{Rotation of the straight line with points $P_1(t)$ and $P_2(t)$ around fixed point $P_3$.}
\end{figure}
\par\noindent
Periodic motion of the points  $P_1(t)$ and $P_2(t)$  on two bound pieces of the curve $X$ on the projective plane  generates  motion by algebraic curves in the phase space, similar to algebraic trajectories in the two fixed centers problem  \cite{eul1,eul2,eul3}.

After canonical transformation of variables
\[u_1 = q_1-\sqrt{q_2},\quad u_2 = q_1+\sqrt{q_2},\quad
p_{u_1}=\dfrac{p_1}{2}+\dfrac{(u_1-u_2)p_2}{2},\quad p_{u_2}=\dfrac{p_1}{2}-\dfrac{(u_1-u_2)p_2}{2}\]
these first integrals  look like
\[\begin{array}{ll}
a=p_1p_2-3q_1^2-q_2\,,\qquad &b=
\dfrac{p_1^2}{4}-q_1p_1p_2+q_2p_2^2+2q_1(q_1^2-q_2)\,,\\ \\
x_3=p_2^2-2q_1\,,\qquad
&y_3=\dfrac{p_1}2+p_2^3-3p_2q_1\,.
\end{array}
\]

Superintegrable Hamiltonian $H=a$ (\ref{wei-int})  belongs to a family of superintegrable Hamiltonians on the plane depending on two integer numbers $k_{1,2}$. Indeed, let us consider Hamiltonian
 \bq\label{wei-int-kk}
H=A_{k_1k_2}=\dfrac{\left(k_1^{-1}p_{u_1}\right)^2}{u_1-u_2}+\dfrac{\left(k_2^{-1}p_{u_2}\right)^2}{u_2-u_1}-u_1^2-u_1u_2-u_2^2\,,
\eq
commuting with the following integral of motion
\[
B_{k_1k_2}=\dfrac{u_2\left(k_1^{-1}p_{u_1}\right)^2}{u_2-u_1}+\dfrac{u_1\left(k_2^{-1}p_{u_2}\right)^2}{u_1-u_2}+(u_1+u_2)u_1u_2\,.
\]
These functions can be obtained from (\ref{wei-int}) by using non-canonical transformation $p_{u_i}\to k_i^{-1}p_{u_i}$, see discussion in \cite{ts18a}. The corresponding quadratures
\[
k_1 \int \dfrac{u_1du_1}{\sqrt{u_1^3+au_1+b}}+
 k_2\int \dfrac{u_2du_2}{\sqrt{u_2^3+au_2+b}}=-2t
\]
and
\[
k_1 \int \dfrac{du_1}{\sqrt{u_1^3+au_1+b}}+ k_2 \int \dfrac{du_2}{\sqrt{u_2^3+au_2+b}}=\mbox{const}\,,
\]
are related to arithmetic equation on the elliptic curve $X$
\[
[k_1]P_1+[k_2]P_2+P_3=0\,.
\]
 Here two  points $[k_1]P_1(t)$ and $[k_2]P_2(t)$ of degree $k_1$ and $k_2$ move along  curve $X$, whereas $P_3$ is a fixed point similar to  Figure 2, but instead of line we have to take curve $Y$ defined by equation $y=\lambda x^{k_1+k_2-1}+\mu ^{k_1+k_2-2}\cdots$.

 According to Euler Hamiltonians $H=A_ {k_1k_2} $ are in  involution with the additional first integral of the form (\ref{eul-int-nm})
 \[
x_3=C_{k_1,k_2}=\left(\dfrac{[k_1]y_1-[k_2]y_2}{[k_1]x_1-[k_2]x_2} \right)^2-\bigl([k_1]x_1+[k_2]x_2\bigr)\,,
\]
which is abscissa $x_3$ of  fixed point $P_3$ that is a well-defined rational function on $x_1,y_1,x_2,y_2$ on the projective plane.

At $k_1=2$ and $k_2=1$ this additional first integral is a rational function of the form
\[\begin{array}{rcl}
C_{21}&=&4   u_1-3   u_2-\dfrac{8   (u_1-u_2)\left(6   u_1^3-12   u_1^2 u_2+6   u_1 u_2^2+p_{u_1} p_{u_2}-2 p_{u_2}^2\right)}{\left(8   u_1^3-12   u_1^2 u_2+4   u_2^3-p_{u_1}^2+4 p_{u_1} p_{u_2}-4 p_{u_2}^2\right)}\\
\\
&+&\dfrac{64  ^2 (u_1-u_2)^4\left(2   u_1^3-3   u_1^2 u_2+  u_2^3+p_{u_1} p_{u_2}-p_{u_2}^2\right)}{\left(8   u_1^3-12   u_1^2 u_2+4   u_2^3-p_{u_1}^2+4 p_{u_1} p_{u_2}-4 p_{u_2}^2\right)^2}\,.
\end{array}
\]
At $k_1=3$ and $k_2=1$ additional first integral is equal to
\[
C_{31}=-(u_1+u_2)+\dfrac{(p_{u_1}^2-9p_{u_2}^2)^2}{81(p_{u_1}+p_{u_2})^2(u_1-u_2)^2}+
\dfrac{8p_{u_1}^2A_1}{9(p_{u_1}+p_{u_2})^2B}-\dfrac{64p_1^5A_2}{9(p_{u_1}+p_{u_2})B^2}
\]
where
\[
B=p_{u_1}^4-8 p_{u_1}^3 p_{u_2}+18\left( p_{u_2}^2- u_1^2 u_2+2 u_1 u_2^2- u_2^3\right) p_{u_1}^2-27\left(p_{u_2}^2-2 u_1^3+3 u_1^2 u_2-u_2^3\right)^2\,,
\]
and
\[\begin{array}{rcl}
A_1&=&
(15 u_1+19 u_2) p_{u_1}^4-6(13 u_1+5 u_2) p_{u_1}^3 p_{u_2} -3 (11 u_1+19 u_2) (2 u_1+u_2) (u_1-u_2)^2 p_{u_1}^2\\ \\
&& +54\bigl((u_1+u_2)p_{u_2}^2+(2u_1+u_2)(u_1-u_2)^3\bigr) p_{u_1}p_{u_2} \\ \\
&&-27 p_{u_2}^2 (5 u_1+u_2) \bigl( p_{u_2}^2-(2u_1+u_2)(u_1-u_2)^2 \bigr)\,,
\\ \\
A_2&=&
2(u_1+u_2) p_{u_1}^4-4 (7 u_1+5 u_2) p_{u_1}^3 p_{u_2}+54 p_{u_2}^2 (5 u_1+u_2)\bigl (p_{u_2}^2-(2u_1+u_2)(u_1-u_2)^2\bigr)\\ \\
&&+3\bigl(24(2 u_1+ u_2) p_{u_2}^2-(7 u_1^2+16 u_1 u_2+13 u_2^2) (u_1-u_2)^2\bigr) p_{u_1}^2\\
\\
&& -9 \bigl(12(3u_1+u_2)p_{u_2}^2-(23u_1^2+38u_1u_2+11u_2^2)(u_1-u_2)^2\bigr) p_{u_1}p_{u_2}\,.
\end{array}
\]
Algebraic trajectories for these superintegrable systems are generated by rotation of a parabola and a cubic around fixed point $P_3$ on the projective plane instead of rotation of the straight line, see the picture in Figure 2.

Similarly we can take superintegrable systems on the plane with Hamiltonians
\[H=p_1p_2+V(q_1,q_2)\]
listed in \cite{ts11,ts08a,ts09} and obtain  families of the superintegrable Hamiltonians $H_{k_1k_2}$  depending on integer numbers $k_{1,2}$, see discussion in \cite{ts18a}.

\subsection{Elliptic coordinates on the plane}
Let us  come back to physical systems and introduce elliptic coordinates on the plane following  Euler \cite{eul1} and Lagrange \cite{lag1,lag2}.  If  $r$ and $r'$ are distances  from a point on the plane  to the two  fixed centers, then elliptic coordinates  $u_{1,2}$ are
\[
r+r'=2u_1\,,\qquad r-r'=2u_2\,.
\]
If two centres are taken to be fixed at $-\kappa$ and $\kappa$ on the  first axis of the Cartesian coordinate system, then we have standard Euler's definition of the elliptic coordinates on the plane
\[q_1 =\dfrac{u_1u_2}{\kappa}\,,\qquad \mbox{and}\qquad q_2 = \dfrac{\sqrt{(u_1^2-\kappa^2)(\kappa^2-u_2^2)}}{\kappa}\,.\]
Coordinates $u_{1,2}$ are curvilinear orthogonal coordinates, which take values only on in the intervals
\[u_2<\kappa<u_1\,,\]
i.e. they are locally defined coordinates.

In terms of  elliptic coordinates $u_{1,2}$ and the corresponding momenta $p_{u_{1,2}}$ kinetic energy has the following form
\[
2T=p_1^2+p_2^2=\dfrac{u_1^2-\kappa^2}{u_1^2-u_2^2}\,p_{u_1}^2
+\dfrac{u_2^2-\kappa^2}{u_2^2-u_1^2}\,p_{u_2}^2\,,
\]
see {\cite{darb,lag2,ser}. By adding separable in elliptic coordinates potentials one gets a Hamiltonian and a second integral of motion
\ben
2H=I_1&=&\dfrac{(u_1^2-\kappa^2)(p_{u_1}^2+V_1(u_1))}{u_1^2-u_2^2}+\dfrac{(u_2^2-\kappa^2)(p_{u_2}^2+V_2(u_1))}{u_2^2-u_1^2}\,,
\nn\\
\label{action-var}\\
I_2&=&-\dfrac{u_2^2(u_1^2-\kappa^2)(p_{u_1}^2+V_1(u_1))}{u_1-u_2}-\dfrac{u_1^2(u_2^2-\kappa^2)(p_{u_2}^2+V_2(u_2))}{u_2-u_1}\,.
\nn
\en
According to Euler and Lagrange there are an equation defining time
\[\dfrac{u_1^2du_1}{\sqrt{(u_1^2-\kappa^2)\Bigl(u_1^2 I_1-V_1(u_1)(u_1^2-\kappa^2)+I_2\Bigr)}}
+\dfrac{u_2^2du_2}{\sqrt{(u_2^2-\kappa^2)\Bigl(u_2^2 I_1-V_2(u_2)(u_2^2-\kappa^2)+I_2\Bigr)}}=dt\,.\]
and an  equation defining trajectories of motion
\[\dfrac{du_1}{\sqrt{(u_1^2-\kappa^2)\Bigl(u_1^2 I_1-V_1(u_1)(u_1^2-\kappa^2)+I_2\Bigr)}}
+\dfrac{du_2}{\sqrt{(u_2^2-\kappa^2)\Bigl(u_2^2 I_1-V_2(u_2)(u_2^2-\kappa^2)+I_2\Bigr)}}=0\,.\]
In these equations $I_{1,2}$ are the values of integrals of motion, see terminology and discussion  in the Lagrange textbook \cite{lag2} and comments by Darboux and Serret  \cite{darb,ser}.

The second equation is reduced to  Euler's differential relation (\ref{eul-eq}) for the Kepler problem
 \bq\label{pot-kepl}
 2H=I_1=p_1^2+p_2^2+\dfrac{\a}{r}\,,\qquad  V_i(u_i)= \dfrac{\a u_i}{u_i^2-\kappa^2} \eq
when
 \[ \int \dfrac{du_1}{\sqrt{(u_1^2-\kappa^2) (I_1u_1^2+\a u_i+I_2)}} + \int \dfrac{du_2}{\sqrt{(u_2^2-\kappa^2) (I_1u_2^2+\a u_2+I_2)}}=const\]
and for the harmonic oscillator
  \bq
  \label{pot-osc}
  2H=I_1=p_1^2+p_2^2-\a^2(q_1^2+q_2^2)\,,\qquad  V_i(u_i)=-\a^2 u_i^2
    \eq
when
\[
 \int \dfrac{du_1}{\sqrt{
(u_1^2-\kappa^2)\bigl(I_1u_1^2+\alpha^2 u_1^2(u_1^2-\kappa^2)+I_2\bigr)
  }}+\int \dfrac{du_2}{\sqrt{
(u_2^2-\kappa^2)\bigl(I_1u_2^2+\alpha^2 u_2^2(u_2^2-\kappa^2)+I_2\bigr)
  }}=const\,.
\]
The equation for the harmonic oscillator  coincides with equation
 \[
 \dfrac{dp}{\sqrt{a_6p^6+a_4p^4+a_2p^2+a_0}}\pm\dfrac{dq}{\sqrt{a_6q^6+a_4q^4+a_2q^2+a_0}}=0\,,
 \]
 which Euler studied in  Problem 82 in \cite{eul0} using reduction of this equation to (\ref{eul-eq}).

Elliptic curve $X$ (\ref{ell-curve}) associated with the Kepler problem  is defined by the following polynomial of fourth order in $x$
\bq\label{ell-c-kepl}
f(x)=I_1x^4-\alpha x^3+(I_2-I_1\kappa^2)x^2+\kappa^2\alpha x-I_2\kappa^2\,,\qquad x_{1,2}=u_{1,2}\,.
\eq
For the harmonic oscillator the corresponding  quartic  polynomial looks like
\bq\label{ell-c-osc}
f(x)=\alpha^2 x^4+(I_1-2\alpha^2\kappa^2)x^3
+(\alpha^2\kappa^4-I_1\kappa^2+I_2)x^2-I_2\kappa^2x\,,\qquad x_{1,2}=u^2_{1,2}\,,
\eq
when abscissas of divisors  $x_{1,2}=u^2_{1,2}$  are equal to the squared  elliptic coordinates.

Substituting elliptic coordinates and first  integrals
 $I_{1,2}$  into $y^2=f(x)$ one gets expressions for ordinates $y_{1,2}$ of points
  $P_1=(x_1,y_1)$ and $P_2=(x_2,y_2)$. For the Kepler problem we have
\[y_1=(u_1^2-\kappa^2)p_{u_1}\qquad\mbox{and}\qquad y_2=(u_2^2-\kappa^2)p_{u_2}\,,
\]
whereas for the harmonic oscillator we obtain
\[y_1=u_1(u_1^2-\kappa^2)p_{u_1}\,,\qquad\mbox{and}\qquad y_2=u_2(u_2^2-\kappa^2)p_{u_2}\,.
\]

Substituting coefficients  $a_4=I_1$, $a_3=-\a$ and  coordinates of divisors  $x_{1,2}$ and  $y_{1,2}$
into the Euler algebraic relation  (\ref{eul-int}) one gets an additional first integral for the Kepler problem
\bq\label{kepl-int}
C=\dfrac{(u_1^2-\kappa^2)(u_2^2-\kappa^2)(p_{u_1}-p_{u_2})^2}{(u_1-u_2)^2}\,,
\eq
 which is independent on first integrals $I_{1,2}$.  This partial integral of motion in the two fixed centers problem and the corresponding algebraic trajectories were studied by  Euler and Lagrange \cite{eul1,eul2,eul3,lag1,lag2}.

Substituting coefficients   $a_4=\a^2$, $a_3=I_1-2\a\kappa^2$ and  coordinates of divisors  $x_{1,2}$ and  $y_{1,2}$
into the Euler algebraic relation  (\ref{eul-int}) one gets an additional first integral for the harmonic oscillator
\bq\label{sw-int}
\begin{array}{rcl}
C&=&
\dfrac{
(u_1^2-\kappa^2)(u_2^2-\kappa^2u_1^2)p_{u_1}^2
-2u_1u_2(u_1^2-\kappa^2)(u_2^2-\kappa^2)p_{u_1}p_{u_2}
+
(u_2^2-\kappa^2)(u_1^2-\kappa^2u_2^2)p_{u_2}^2
}{(u_1^2-u_2^2)^2}\\
\\
&+&\displaystyle\a^2\kappa^2(u_1^2+u_2^2)\,.
\end{array}
\eq
First integrals  (\ref{kepl-int}) and  (\ref{sw-int}) have two different forms in elliptic and especially in Cartesian coordinates, but they have a common simple form in terms of  coordinates of divisors  (\ref{eul-int}).

Other superintegrable systems separable in elliptic coordinates on the plane with first integral of the form
(\ref{eul-int})  are discussed in \cite{km15,ts09a,ts09,ts10,ts12}.

\subsection{Elliptic coordinates on the sphere}
Two-dimensional metrics which geodesic flows admit three functionally independent  first integrals are called
superintegrable metrics. Superintegrable metrics with first integrals which are second order polynomials in momenta  were described by Koenigs \cite{koen}.  In this Section we consider a well-known  superintegrable metric on the two-dimensional sphere with quadratic first integrals, one of which has the Euler form (\ref{eul-int}). In the next Section we present superintegrable metrics on the sphere with algebraic and rational first integrals which are easily constructed using multiplication of points on elliptic curve. We have to underline that our main aim is a search of algebraic trajectories for dynamical systems following to Euler \cite{eul1,eul2,eul3} and Lagrange \cite{lag1,lag2}. Construction of the first integrals is only a  tool for the searching of such trajectories.

Let us  introduce elliptic coordinates on the two-dimensional sphere  $\mathbb S^2\subset\mathbb R^3$ embedded into three-dimensional Euclidean space with Cartesian coordinates $q_1,q_2$ and $q_3$
The elliptic coordinate system $u_{1,2}$ on the sphere $\mathbb S^2$ with parameters $\a_1<\a_2<\a_3$ is defined through  equation
\[
\dfrac{q_1^2}{z-\a_1}+\dfrac{q_2^2}{z-\a_2}+\dfrac{q_3^2}{z-\a_3}=\dfrac{(z-u_1)(z-u_2)}{\phi(z)}\,,\qquad \phi(z)=(z-\a_1)(z-\a_2)(z-\a_3)\,,
\]
which should be interpreted as an identity with respect to $z$. It implies
\[
q_1^2+q_2^2+q_3^2=1\,,
\]
which is a standard description of the sphere in $\mathbb R^3$.  Similar to  elliptic coordinates on the plane,  elliptic coordinates on the sphere  are also orthogonal and only locally defined. They take values in the intervals
\[
\a_1<u_1<\a_2<u_2<\a_3\,.
\]
The coordinates and the parameters can be subjected to a simultaneous linear
transformation $u_i\to au_i+b$ and $\a_i\to a\a_i+b$, so it is always possible to choose
$\a_1=0$ and $\a_3=1$.

Let us consider  free motion on the sphere $\mathbb S^2$ defined by the Hamiltonian
\[
H=p_1^2+p_2^2+p_3^2
\]
In  elliptic coordinates this Hamiltonian and the corresponding second integral of motion have the following form
\bq\label{sphere-ham}
2H=I_1=\dfrac{\phi(u_1)p_{u_1}^2}{u_1-u_2}+\dfrac{\phi(u_2)p_{u_2}^2}{u_2-u_1}\,,\qquad
I_2=\dfrac{u_2\phi(u_1)p_{u_1}^2}{u_2-u_1}+\dfrac{u_1\phi(u_2)p_{u_2}^2}{u_1-u_2}\,.
\eq
As above, there are two Abel's equations
\[
\begin{array}{rcl}
 \dfrac{u_1du_1}{\sqrt{\phi(u_1)(u_1I_1+I_2)}}+\dfrac{u_2du_2}{\sqrt{\phi(u_2)(u_2I_1+I_2)}}&=&2dt\,,\\ \\
  \dfrac{du_1}{\sqrt{\phi(u_1)(u_1I_1+I_2)}}+\dfrac{du_2}{\sqrt{\phi(u_2)(u_2I_1+I_2)}}&=&0\,.
\end{array}
\]
defining time and trajectories according to Euler and Lagrange terminology \cite{lag2}.

It is easy to see that second equation
\[
 \dfrac{du_1}{\sqrt{f(u_1)}}+\dfrac{du_2}{\sqrt{f(u_2)}}=0\,,
\]
coincides with Euler equation (\ref{eul-eq}) on the elliptic curve $X$ defined by
\[
y^2=f(x)\,,\qquad  f(x)=\phi(x)(xI_1+I_2)=a_4x^4+a_3x^3+a_2x^2+a_1x+a_0\,.
\]
It allows us to directly obtain an additional integral of motion  (\ref{eul-int}) which has the following form in elliptic coordinates
\bq\label{sphere-int}
\begin{array}{rcl}
C&=&-\dfrac{
\bigl(\a_1\a_2\a_3-(\a_1 \a_2 +\a_1 \a_3 +\a_2 \a_3) u_1+(\a_1+\a_2 +\a_3) u_2^2-u_1 u_2^2\bigr) \phi_1p_{u_1}^2}{(u_1-u_2)^2}\\ \\
&&-
\dfrac{
\bigl(\a_1\a_2\a_3-(\a_1 \a_2 +\a_1 \a_3 +\a_2 \a_3) u_2+(\a_1 +\a_2 +\a_3) u_1^2-u_1^2 u_2\bigr)\phi_2p_{u_2}^2}{(u_1-u_2)^2}\\ \\
&&-\dfrac{ 2\phi_1\phi_2\,p_{u_1}\,p_{u_2} }{(u_1-u_2)^2}\,.
\end{array}
\eq
 In modern terms, two-dimensional metric $\mathrm g(u_1,u_2)$ in (\ref{sphere-ham})
\[H=\sum {\mathrm g}_{ij}p_{u_i}p_{u_j}\,,\qquad
\mathrm g=\left(
                           \begin{array}{cc}
                             \frac{(u_1-\a_1)(u_1-\a_2)(u_1-\a_3)}{u_1-u_2} & 0 \\
                             0 & \frac{(u_2-\a_1)(u_2-\a_2)(u_2-\a_3)}{u_2-u_1} \\
                           \end{array}
                         \right)
\]
is  a superintegrable metric on the  sphere, thus, we have global superintegrable Hamiltonian
system on a  compact manifold with closed trajectories. Recall, that  problem of finding and describing
global integrable Hamiltonian systems on a compact manifold is one of the central
topics in the classical mechanics, see discussion in  \cite{bol95}.

Summing up, equations of motion  for the Kepler system, for the harmonic oscillator  on the plane and for the geodesic motion on the sphere are reduced to equation which coincide with Abel's equation defining rotation of the parabola around a fixed point on the elliptic curve, see Figure 3.
\begin{figure}[H]
\center{\includegraphics[width=0.7\linewidth, height=0.2\textheight]{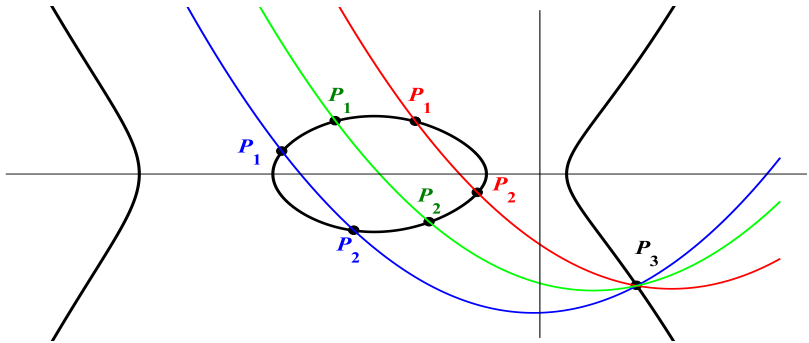} }
\caption{Rotation of the parabola with points $P_1(t)$ and $P_2(t)$ around fixed point $P_3$.}
\end{figure}
\par\noindent
For all these superintegrable systems motion of two points  $P_1(t)$ and $P_2(t)$  on  bound pieces of the curve $X$  generates  motion by algebraic trajectories in the phase space $T^*\mathbf R^2$ or $T^8\mathbf S^2$.

For all these superintegrable systems additional first integrals (\ref{kepl-int},\ref{sw-int}) and (\ref{sphere-int}) are defined by coordinates
of this fixed point $P_3$ by equation (\ref{eul-int-4}). Of course, abscissa $x_3$ and ordinate $y_3$ are also first integrals depending on $I_1,I_2$ and $C$.

\section{Superintegrable systems with algebraic and rational first integrals}
In this Section we consider Abel's equations (\ref{eul-eq-nm})
\[
k_1\dfrac{dx_1}{y_1}+k_2\dfrac{dx_2}{y_2}=0\,,
\]
defining motion of the straight line, quadric, cubic, quartic and so on  around a fixed point on the elliptic curve when
two movable points $P_{1}(t)$ and $P_2(t)$ of degree $k_{1,2}$ and one fixed point $P_3$ form an intersection divisor of elliptic curve $X$ with the straight line, quadric, cubic, quartic, etc.

 In order to construct superintegrable systems associated with this motion of points on the elliptic curve  we just have to identify
 Abel's equations (\ref{eul-eq-nm}) on a projective plane with Abel's equations on some phase space.

\subsection{New superintegrable systems on the plane}
Let us consider integrable systems  with the following Hamiltonian and second integral of motion
\ben
2H=I_1&=&\dfrac{(u_1^2-\kappa^2)\left(\left(\frac{p_{u_1}}{k_1}\right)^2+V_1\right)}{u_1^2-u_2^2}
+\dfrac{(u_2^2-\kappa^2)\left(\left(\frac{p_{u_2}}{k_1}\right)^2+V_2\right)}{u_2^2-u_1^2}\,,
\nn\\
\label{plane-ham}\\
I_2&=&-\dfrac{u_2^2(u_1^2-\kappa^2)\left(\left(\frac{p_{u_1}}{k_1}\right)^2+V_1\right)}{u_1-u_2}
-\dfrac{u_1^2(u_2^2-\kappa^2)\left(\left(\frac{p_{u_2}}{k_1}\right)^2+V_2\right)}{u_2-u_1}\,,
\nn
\en
where $u_{1,2}$ and $p_{u_{1,2}}$ are elliptic coordinates and  the corresponding momenta. At $k_1=k_2$ these integrals of motion coincide with the integral of motion (\ref{action-var}) up to the scalar factor.

If potentials $V_{1,2}$ are given by (\ref{pot-kepl})  or  (\ref{pot-osc}),  and $k_{1,2}$ are integer positive numbers, then all the trajectories of motion are algebraic  trajectories because equation defined trajectories coincides with the  Abel's equation
\[
k_1\dfrac{dx_1}{y_1}+k_2\dfrac{dx_2}{y_2}=0\,,\qquad k_{1,2}\in\mathbb Z_+\,,
\]
on the elliptic curve
\[
y^2=f(x)\,,\qquad  f(x)=a_4x^4+a_3x^3+a_2x^2+a_1x+a_0\,.
\]
For the Kepler potential and the harmonic oscillator potential quartic polynomials $f(x)$ are given by (\ref{ell-c-kepl}) and (\ref{ell-c-osc}), respectively. Consequently,  complete integral $C_{k_1k_2}$ (\ref{eul-int-nm}) of the Abel's equation  gives rise to a complete first integral for the corresponding Hamiltonian system.

\begin{prop}
 Hamiltonian $H=2I_1$ (\ref{plane-ham}) commutes with integral of motion  $I_2$
\[\{H,I_2\}=0\,.\]
for arbitrary potentials $V_{1,2}$ and parameters $k_{1,2}$.

If  potentials are equal to
\[
 V_i(u_i)= \dfrac{\a u_i}{u_i^2-\kappa^2}\quad\mbox{or}\quad
  V_i(u_i)=-\a^2 u_i^2\,,
\]
and $k_{1,2}$ are positive integers, then Hamiltonian $H=2I_1$ (\ref{plane-ham})  commutes
\[
\{H,C_{k_1,k_2}\}=0\,.
\]
 with additional first integral $C_{k_1,k_2}$ (\ref{eul-int-nm}) which is independent on first integrals $I_{1,2}$.
 \end{prop}
 The proof of this proposition is completely based on the Euler solution of Problem 83 in \cite{eul0}.

 For the Kepler potential we have to substitute into $C_{k_1,k_2}$ (\ref{eul-int-nm}) the following   coordinates of divisors
 \[
 x_{i}=u_{i}\,,\qquad  y_i=(u_i^2-\kappa^2)\frac{p_{u_i}}{k_i}\,,\qquad i=1,2,
 \]
 and coefficients
 \[
a_4=I_1,\quad a_3=\a,\quad a_2=I_2-\kappa^2I_1,\quad
a_1=\a\kappa^2,\quad a_0=-\kappa^2I_2\,,
\]
  whereas for the harmonic oscillator potential these coordinates  and coefficients are equal to
 \[
  x_{i}=u_{i}^2\,,\qquad  y_i=u_i(u_i^2-\kappa^2)\frac{p_{u_i}}{k_i}\,,\qquad i=1,2,
 \]
and
\[
a_4=\alpha^2,\quad a_3=I_1-2\alpha^2\kappa^2,\quad a_2=\alpha^2\kappa^4-\kappa^2I_1+I_2,\quad
a_1=-\kappa^2I_2,\quad a_0=0\,.
\]
We calculated integrals $C_{21}$ and $C_{31}$  in two computer algebra systems Mathematica and Maple and
directly verified that these integrals are in involution  with the corresponding Hamiltonians  $H=2I_1$ (\ref{plane-ham}).

In case of the Kepler potential (\ref{pot-kepl}) additional first integrals $C_{21}$ ({\ref{eul-int-21}}) depends on $\sqrt{a_4}=\sqrt{I_1}$,
i.e. it is the algebraic function on momenta $p_{u_{1,2}}$.   Additional first integral   $C_{31}$  ({\ref{eul-int-31}})  is  the rational function on elliptic coordinates $u_{1,2}$ and momenta $p_{u_{1,2}}$.

In  case of the harmonic oscillator potential (\ref{pot-osc}) both the additional integrals of motion  $C_{21}$ and  $C_{31} $ are rational functions on elliptic coordinates $u_{1,2}$ and momenta $p_{u_{1,2}}$. For instance, we explicitly present additional first integral $C_{21}$, which is in  involution with the Hamiltonian
\[
2H=I_1=\dfrac{ (u_1^2-\kappa^2)p_{u_1^2}}{4(u_1^2-u_2^2)}+
\dfrac{(\kappa^2-u_2^2)p_{u_2}}{u_1^2-u_2^2}+\a^2(\kappa^2-u_1^2-u_2^2)\,,
\]
and does not commute with the second polynomial integral of motion
\[
I_2=\dfrac{u_2^2(\kappa^2-u_1^2)p_{u_1}^2}{4(u_1^2-u_2^2)}
-\dfrac{u_1^2(\kappa^2-u_2^2)p_{u_2}^2}{u_1^2-u_2^2}+\alpha^2u_1^2u_2^2\,.
\]
Using various tools in Mathematica and Maple one gets the following  observable expression
\[
C_{21}=-\kappa^2H+(u_2^2-\kappa^2)(2\alpha u_2+p_{u_2})p_{u_2}+\alpha^2(\kappa^4-\kappa^2 u_2^2+u_2^4)
+\dfrac{p_{u_1}C_1}{D}+\dfrac{p_{u_1}^2C_2}{D^2}\,,
\]
where
\[
D=\dfrac{(\kappa^2u_2^2+u_1^4-2u_1^2u_2^2)p_{u_1}^2}{\kappa^2-u_2^2}
+\dfrac{4\alpha u_1(u_1^2-u_2^2)^2(\alpha u_1+p_{u_1})}{\kappa^2-u_2^2}-4u_1u_2p_{u_1}p_{u_2}+4u_1^2p_{u_2}^2\,,
\]
and
\[\begin{array}{rcl}
C_1&=&\scriptstyle u_2^2 \left(2 \kappa^2-u_1^2-2 u_2^2\right) p_{u_1}^3
+2 u_1 u_2 \bigl(\alpha u_2( \kappa^2- u_1^2)+( u_1^2+4  u_2^2-3 \kappa^2 )p_{u_2}\bigr) p_{u_1}^2\\
\\
&-&\scriptstyle 4\Bigl(
\alpha^2u_2^2(u_1^2-u_2^2)(\kappa^2-u_1^2-u_2^2)
+2\alpha (\kappa^2u_1^2-2\kappa^2u_2^2-u_1^4+2u_2^4)u_2p_{u_2}+\bigl(\kappa^4-(u_1^2+8u_2^2)\kappa^2+u_2^2(2u_1^2+7u_2^2)\bigr)p_{u_2}^2
\Bigr) p_{u_1}\\
\\
&+&\scriptstyle 8 u_1 (\alpha  u_2+p_{u_2}) \bigl(\alpha (u_1^2- u_2^2)-p_{u_2} u_2\bigr)
 \bigl(\alpha u_2(u_1^2- u_2^2)+(\kappa^2-u_2^2) p_{u_2}\bigr)
\end{array}
\]
\[
\begin{array}{rcl}
C_2&=&\scriptstyle u_2^2\bigl(4u_1^2u_2^2-\kappa^2(u_1^2+2u_2^2)\bigr) p_{u_1}^4-4u_1u_2
\Bigl(\a u_2\left(\kappa^2u_1^2-3u_1^2u_2^2+2u_2^4\right)-\bigl(\kappa^2(u_1^2+4u_2^2)-5u_1^2u_2^2-2u_2^4\bigr)p_{u_2}\Bigr)p_{u_1}^3\\
\\
&-&\scriptstyle  4u_2\Bigl(
\alpha^2\kappa^2u_2(u_1^2-u_2^2)^2
-4\a\left(\kappa^2(u_1^4+u_1^2u_2^2-u_2^4)-5u_1^4u_2^2+4u_1^2u_2^4\right)p_{u_2}
-u_2\left(\kappa^4-\kappa^2(12u_1^2-7u_2^2)+24u_1^2u_2^2\right)p_{u_2}^2
\Bigr)p_{u_1}^2\\  \\
&-&\scriptstyle 16u_1u_2\Bigl(
\alpha^2(u_1^2-u_2^2)^2\left(\alpha u_2^3-(\kappa^2-5u_2^2)p_{u_2}\right)
+\a u_2\left(\kappa^2(6u_1^2-5u_2^2)-12u_1^2u_2^2+11u_2^4\right)p_{u_2}^2\Bigr.\\
&&\scriptstyle\qquad\qquad\qquad\qquad\qquad\qquad\qquad\qquad\qquad\qquad\qquad \Bigl.+\left(\kappa^4-2\kappa^2(u_1^2-+4u_2^2)+4u_1^2u_2^2+7u_2^4\right)p_{u_2}^3
\Bigr)p_{u_1}\\
\\
&+&\scriptstyle 16u_1^2
\Bigl(
4\a u_2(u_1^2-u_2^2)(\kappa^2-2u_2^2)+(\kappa^4-8\kappa^2u_2^2+8u_2^4)p_{u_2}
\Bigr)
p_{u_2}^3\,.
\end{array}
\]
 In our opinion, it is practically  impossible to use such sophisticated  expressions in the direct search of additional integrals of motion or for investigations of algebras of integrals of motion, see also discussion in \cite{km15}. Nevertheless, because additional first integrals of motion are easily expressed via coordinates of third point $P_3=[k_1]P_1+[k_2]P_2$ at any $k_{1,2}$, for instance
\[C_{k_1k_2}=2a_4x_3^2+a_3x_3+a_2-2\sqrt{a_4}y_3\,, \]
we can derive the algebra of integrals using  the well-known syzygies on elliptic curve \cite{gr,hens}.

\subsection{New superintegrable metrics on the sphere}
Let us consider geodesic motion on the two-dimensional sphere $\mathbb S^2$ defined by Hamiltonian and second integral of motion
 \bq\label{sphere-hamk}
H=I_1=\dfrac{\phi(u_1)p_{u_1}^2}{k_1^2(u_1-u_2)}+\dfrac{\phi(u_2)p_{u_2}^2}{k_2^2(u_2-u_1)}\,,\qquad
I_2=\dfrac{u_2\phi(u_1)p_{u_1}^2}{k_1^2(u_2-u_1)}+\dfrac{u_1\phi(u_2)p_{u_2}^2}{k_2^2(u_1-u_2)}\,.
\eq
At $k_1=k_2$ these first integrals coincide with (\ref{sphere-ham}) up to the scalar factor $k_1^2$.

Trajectories of motion are defined as solutions of equation
\[
\begin{array}{rcl}
  k_1\dfrac{du_1}{\sqrt{\phi(u_1)(u_1I_1+I_2)}}+k_2\dfrac{du_2}{\sqrt{\phi(u_2)(u_2I_1+I_2)}}&=&0\,,
\end{array}
\]
which is the Abel type equation (\ref{eul-eq-nm})
\[
k_1\dfrac{dx_1}{y_1}+k_2\dfrac{dx_2}{y_2}=0\,,\qquad x_{1,2}=u_{1,2}\,,
\]
on the elliptic curve defined by an equation of the form
\[
y^2=f(x)\,,\qquad  f(x)=\phi(x)(xI_1+I_2)=a_4x^4+a_3x^3+a_2x^2+a_1x+a_0\,.
\]
Consequently, we have superintegrable systems with additional first integral $C_{k_1k_2}$ (\ref{eul-int-nm}).

\begin{prop}
The following Hamiltonian $H=\sum \mathrm g_{ij}p_{u_i}p_{u_j}$ on the two-dimensional sphere with diagonal metric
\[
\mathrm g=\dfrac{1}{u_1-u_2}\left(
                           \begin{array}{cc}
                             \dfrac{(u_1-\a_1)(u_1-\a_2)(u_1-\a_3)}{k_1^2} & 0 \\
                             0 & \dfrac{(\a_1-u_2)(\a_2-u_2)(\a_3-u_2)}{k_2^2} \\
                           \end{array}
                         \right)\,,\qquad k_1,k_2\in\mathbb Z_+
\]
is in  involution with two independent first integrals
\[
\{H,I_2\}=0\,,\qquad \{H,C_{k_1,k_2}\}=0\,.
\]
Thus, this metric is a superintegrable metric.
\end{prop}
 The proof of this proposition is completely based on the Euler solution of  Problem 83 in \cite{eul0}.

In Cartesian  variables $q_1,q_2,q_3$ and momenta $p_1,p_2,p_3$ in $T^*\mathbb R^3$, so that
\[
q_1^2+q_2^2+q_3^2=1\,,\qquad q_1p_1+q_2p_2+q_3p_3=0\,,
\]
 this Hamiltonian has the following form
\[\begin{array}{rcl}
H&=&
 \dfrac{k_1^2+k_2^2}{8k_1^2k_2^2}\,(p_1^2+p_2^2+p_3^2)\\ \\
&+&\dfrac{k_1^2-k_2^2}{2k_1^2k_2^2(u_2-u_1)}\left(
\a_1q_2q_3p_2p_3+\a_2q_1q_3p_1p_3+\a_3q_1q_2p_1p_2+\b_1p_1^2+\b_2p_2^2+\b_3p_3^2\right)\,,
\end{array}
\]
where
\[
\begin{array}{rcl}
\b_1&=&\frac{1}{4}\left( (\phantom{-}q_2^2+q_3^2)\a_1+(\phantom{-}q_1^2-q_3^2)\a_2+(\phantom{-}q_1^2-q_2^2)\a_3 \right)\,,\\ \\
\b_2&=&\frac{1}{4}\left( (\phantom{-}q_2^2-q_3^2)\a_1+(\phantom{-}q_1^2+q_3^2)\a_2+(-q_1^2+q_2^2)\a_3 \right)\,,\\ \\
\b_3&=&\frac{1}{4}\left( (-q_2^2+q_3^2)\a_1+(-q_1^2+q_3^2)\a_2+(\phantom{-}q_1^2+q_2^2)\a_3\right)\,,
\end{array}
\]
and
\[\begin{array}{rcl}
{u_2-u_1}&=&\Bigl(
(\a_3-\a_2)^2 q_1^4+(\a_1-\a_3)^2 q_2^4+(\a_2-\a_1)^2 q_3^4+2 (\a_3-\a_2)(\a_3-\a_1) q_1^2q_2^2\Bigr.
\\
&&\Bigl.
+2 (\a_3-\a_2) (\a_1-\a_2) q_1^2q_3^2+ 2 (\a_3-\a_1) (\a_2-\a_1) q_2^2 q_3^2\Bigr)^{1/2}
\end{array}
\]
is a difference of the elliptic coordinates, which is the globally defined strictly positive function on the sphere.

At $k_1=2,3$ and  $k_2=1$ explicit expression for the first integral $C_{k_1,k_2}$ can be obtained substituting
\[
x_{1,2}=u_{1,2}\,,\qquad y_{1,2}=(u_{1,2}-\a_1)(u_{1,2}-\a_2)(u_{1,2}-\a_3)\,\frac{p_{u_{1,2}}}{k_{1,2}}
\]
and
\[\begin{array}{ll}
a_4=I_1,\quad a_3=I_2-(\a_1+\a_2+\a_3)I_1\,\quad
&a_2=(\a_1\a_2+\a_1\a_3+\a_2\a_3)I_1-(\a_1+\a_2+\a_3)I_2\,\\ \\
a_1=(\a_1\a_2+\a_1\a_3+\a_2\a_3)I_2-\a_1\a_2\a_3I_1\,,\quad
&a_0=-\a_1\a_2\a_3I_2
\end{array}
\]
into the divisors doubling and tripling operations  (\ref{doub-ell},\ref{trip-ell}) and then into the definition of Euler's integral (\ref{eul-int-nm}).

Because $\sqrt{a_4}=\sqrt{I_1} $,  first integral $C_{21}$ (\ref{eul-int-21}) is an algebraic function  on the elliptic coordinates $u_{1,2}$ and the corresponding momenta $p_{u_{1,2}}$. Additional first integral $C_{31}$  is a rational function on variables of separation of the form (\ref{eul-int-31}).

As above at $k_1=2,3$ and  $k_2=1$ we directly verified that Hamiltonians $H=\sum \mathrm g_{ij}p_{u_i}p_{u_j}$ and first integrals $C_{k_1,k_2}$ are in  involution by using computer algebra systems Mathematica and Maple.

\section{Conclusion}
Consider motion of  $k$ points $P_1,\ldots, P_k$ around $m$ fixed points $P_{k+1},\ldots, P_{k+m}$ along a plane curve $X$, which is governed by  Abel's equations generated by the addition of points  on $X$
\[
\left(P_1+\cdots+P_k\right) + \left(P_{k+1}+\cdots+P_{k+m}\right)=0\,.
\]
If the same Abel's equations arise when studying  motion of an integrable Hamiltonian or non-Ha\-mil\-to\-ni\-an system, then this dynamical system  is a superintegrable system with additional partial or complete integrals of motion which are given by the coordinates of   fixed points $P_{k+1},\ldots, P_{k+m}$. According to Abel's theorem  these integrals are algebraic functions in coordinates  of movable points $P_{1},\ldots, P_{m}$ and, therefore, they are well-defined algebraic functions on original physical variables. Evolution of  movable points around the fixed points  gives rise to algebraic trajectories of this superintegrable system similar to algebraic trajectories in the Euler two centers problem.

In this note we study motion of two points of degree $k_1$ and $k_2$ around one fixed point on the elliptic curve.  The corresponding Abel's equations also arise when studying  the following superintegrable systems with two degrees of freedom: the Kepler problem, the harmonic oscillator, the geodesic motion on the sphere and their superintegrable deformations. Here we only present the corresponding first integrals, which are polynomial functions in momenta at $k_1=k_2$  and  algebraic/rational functions at $k_1\neq k_2$. We plan to  discuss the corresponding algebraic trajectories in forthcoming publication.

Arithmetic on elliptic curves has  been  an  object  of  study  in  mathematics  for  well  over  a century.
Recently arithmetic on elliptic curves  has  proven  useful  in  applications  such  as factoring \cite{leh87}, elliptic curve cryptography \cite{blake99,hand06,was08}, and in the proof of Fermat’s last theorem \cite{wil95}. In real world elliptic curve point multiplication is one of the most widely used methods for digital signature schemes in cryptocurrencies, which  is applied in both Bitcoin and Ethereum for signing transactions.   In \cite{ts17a,ts17b,ts18b} we apply elliptic and hyperelliptic curve point multiplication to discretization of some known integrable systems in Hamiltonian and non-Hamiltonian mechanics and to construction of new integrable Hamiltonian systems in \cite{ts17b,ts17c} . In this note we use this universal mathematical tool to construct  new superintegrable systems with algebraic and rational integrals of motion. It will be interesting to discuss other possible  applications of arithmetic on elliptic and hyperelliptic curves in classical and quantum mechanics.

The work was supported by the Russian Science Foundation  (project  18-11-00032).

\end{document}